\pdfoutput=1

\documentclass[11pt]{article}

\usepackage[final]{acl}

\usepackage{times}
\usepackage{latexsym}
\usepackage{listings}
\usepackage{lstautogobble}
\usepackage[T1]{fontenc}

\usepackage[utf8]{inputenc}

\usepackage{microtype}

\usepackage{inconsolata}

\usepackage{graphicx}

\usepackage{array}
\usepackage{pifont}
\usepackage{tabularx}
\usepackage{adjustbox}
\usepackage{multirow}
\usepackage{enumitem}
\usepackage{xspace}
\usepackage{tcolorbox}
\usepackage{booktabs,amsfonts,dcolumn}
\usepackage{hyperref}
\usepackage{url}
\usepackage{amsmath,amsthm,amsfonts,amssymb,bm,stmaryrd,bbm}
\usepackage[noorphans,vskip=0.75ex,leftmargin=2ex]{quoting}
\usepackage{footmisc}\interfootnotelinepenalty=10000
\usepackage{colortbl}
\usepackage{cleveref}

\newcommand\tf[1]{\textbf{#1}}
\newcommand\ttt[1]{\texttt{#1}}

\newcommand{\ours}{LitSearch}

\newcommand{\instructor}{Instructor}
\newcommand{\gtr}{GTR}
\newcommand{\efive}{E5}
\newcommand{\grit}{GritLM}
\newcommand{\instructorfull}{Instructor-XL}
\newcommand{\gtrfull}{GTR-T5-large}
\newcommand{\efivefull}{E5-large-v2}
\newcommand{\gritfull}{GritLM-7B}

\newcommand{\inlineq}{inline-citation}
\newcommand{\authorq}{author-written}
\newcommand{\inlineQ}{Inline-citation}
\newcommand{\authorQ}{Author-written}
\newcommand{\Inlineq}{Inline-Citation}
\newcommand{\Authorq}{Author-written}

\title{Citation Benchmark}
\title{\ours{}: A Retrieval Benchmark for Scientific Literature Search}

\author{Anirudh Ajith$^1$\quad Mengzhou Xia$^1$\quad Alexis Chevalier$^{1,2}$\quad Tanya Goyal$^1$\\\tf{Danqi Chen}$^1$\quad \tf{Tianyu Gao}$^1$\\
$^1$Princeton Language and Intelligence (PLI), 
Princeton University\quad $^2$BCG X\\
\ttt{\{anirudh.ajith,mengzhou,achevalier,tanyagoyal,}\\
\ttt{danqic,tianyug\}@princeton.edu}
}

\begin{document}
\maketitle

\begin{abstract}

Literature search questions, such as ``{Where can I find research on the evaluation of consistency in generated summaries?}'' pose significant challenges for modern search engines and retrieval systems. These questions often require a deep understanding of research concepts and the ability to reason across entire articles. In this work, we introduce \emph{\ours{}}, a retrieval benchmark comprising 597 realistic literature search queries about recent ML and NLP papers. \ours{} is constructed using a combination of (1) questions generated by GPT-4 based on paragraphs containing inline citations from research papers and 
(2)  questions manually written by authors about their recently published papers. 
All \ours{} questions were manually examined or edited by experts to ensure high quality. We extensively benchmark state-of-the-art retrieval models and also evaluate two LLM-based reranking pipelines.
We find a significant performance gap between BM25 and state-of-the-art dense retrievers, with a 24.8\% absolute difference in recall@5. The LLM-based reranking strategies further improve the best-performing dense retriever by 4.4\%. Additionally, commercial search engines and research tools like Google Search perform poorly on \ours{}, lagging behind the best dense retriever by up to 32 recall points. Taken together, these results show that \ours{} is an informative new testbed for retrieval systems while catering to a real-world use case.\footnote{Our dataset and code are available at \url{https://github.com/princeton-nlp/LitSearch}.}

\end{abstract}

\section{Introduction}
\label{sec:intro}

Finding literature via a specific search query---for example, collecting related work, checking if a method has been proposed before, or recalling a previously seen paper---is a critical task for researchers. 
Developing systems that recommend citations pertinent to such inquiries holds the potential to enhance researchers' productivity and expedite scientific discovery~\cite{farber2020citation}.
However, this task is inherently challenging as it often requires deep domain expertise and reasoning through lengthy papers. %

\begin{figure}[t]
    \centering
    \includegraphics[width=0.98\linewidth]{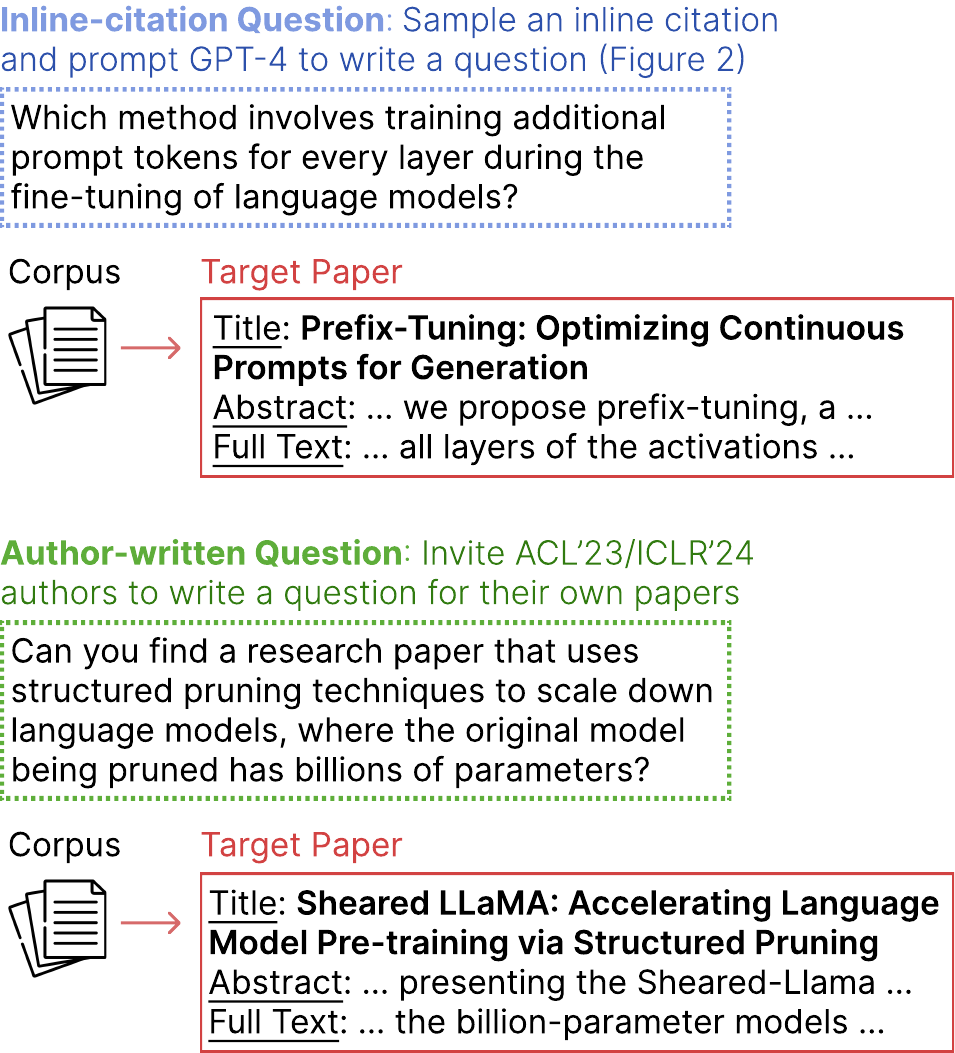}
    \caption{Examples of \textit{\inlineq{}} and \textit{\authorq{} questions} from  \ours{}. These questions are often challenging and require a deep understanding of the target papers to answer correctly. 
    }
    \label{fig:teaser}
\end{figure}

\begin{figure*}[t]
    \centering
    \includegraphics[width=0.98\textwidth]{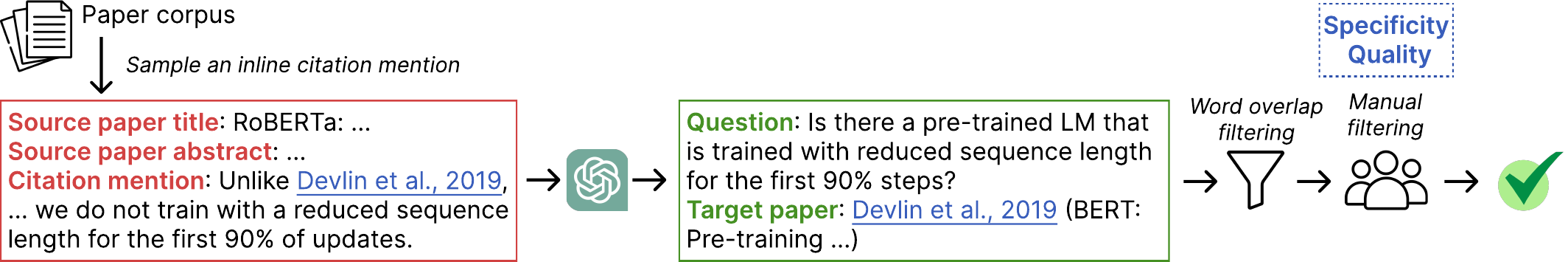}
    \caption{The pipeline for generating \inlineq{} questions. We first sample a citation mention and prompt GPT-4 to generate a question. Next, we filter questions based on word overlap with the target paper title and perform manual inspections to annotate their specificity and quality (see rubrics in \Cref{tab:manual_filter}). 
    }
    \label{fig:inline}
\end{figure*}

Prior to this study, the task of citation recommendation was 
often formalized by 
using inline citation mentions from existing papers as queries, and the cited papers as targets~\cite{he2010context, gu2022local}.
For instance, given the citation mention ``\textit{RoBERTa and T5 are based on recent advances in masked language modeling \ttt{[citation]}},'' the text surrounding the citation mention is used as a retrieval query, and the cited paper is the target literature. 
However, directly using inline citations often leads to queries that are noisy, overly broad (e.g., ``\textit{Large Language Models \ttt{[citation]}}''), or highly context-dependent (e.g., ``\textit{We follow the hyperparameters of \ttt{[citation]}}'').

In this work, we propose a new literature retrieval benchmark called \emph{\ours{}}. As illustrated in \Cref{fig:teaser}, a literature search question seeks papers that meet specific criteria, closely reflecting actual research workflows.
\ours{} consists of two subsets:
(1) For \tf{\inlineq{} questions}, we sample citation mentions from a collection of scientific papers and use GPT-4~\citep{openai2023gpt4} to rewrite them into literature search questions (\Cref{fig:inline}). 
We retain questions with a low word overlap with the title of the target papers, and perform manual examination to ensure high quality. 
(2) For \tf{\authorq{} questions}, we invited authors of ACL 2023 and ICLR 2024 papers to write literature search questions for their own papers. 
This subset is also manually examined and filtered to remove any inaccurate or easy questions.
\ours{} contains 597 questions in total, each paired with one or more scientific papers as the ground truth. 

\ours{} has several unique characteristics: 
(1) To the best of our knowledge, \ours{} is the \emph{first} dataset featuring realistic literature search questions, providing a new testbed for citation recommendation and retrieval systems.
(2) \ours{} is \emph{challenging}, requiring deep understanding and reasoning over entire articles. The average document length (6,041/134 words for full texts/titles and abstracts) is significantly longer than that of most existing retrieval benchmarks (e.g., 56 for \citep{nguyen2017ms}).
(3) \ours{} is of \emph{high quality}, with all questions manually examined by the authors.

We conduct extensive experiments 
on both state-of-the-art retrieval models and 
reranking with large language models (LLMs). 
On \ours{}, the best dense retrieval model, \grit{}~\citep{muennighoff2024generative}, achieves an average recall@5
of 74.8\%, outperforming BM25~\citep{bm25} by 24.8\%.
The recall@5 of \grit{} is further improved by 4.4\% with GPT-4o reranking.
On the other hand, commercial search engines and research tools like Google Search perform poorly on this task, only achieving an average recall@5 of 42.8\% at most. %
Furthermore, compared to existing retrieval datasets from %
BEIR~\citep{thakur2021beir} and MTEB~\citep{muennighoff2022mteb}, 
\ours{} effectively reflects the performance differences among various embedding models, 
making it an informative testbed for evaluating state-of-the-art retrieval systems.

\section{\ours{}}

Our benchmark \ours{} consists of 
(a) a large corpus of scientific papers $\mathcal{P}$ and 
(b) pairs of literature search questions and one or more target papers from $\mathcal{P}$. %
Our desiderata are scientific questions that researchers may use while conducting literature surveys. 
We use two different strategies to collect such questions: (1) we construct questions using the surrounding context from inline citations in published papers (\Cref{sec:inline}), and (2) we invited the authors of recent conference publications to manually write questions about their own papers (\Cref{sec:author}). For both subsets, we ensure high question quality via manual inspection and filtering conducted by the authors of this work (\Cref{sec:filter}).

\subsection{\inlineQ{} Questions}
\label{sec:inline}
We define the following concepts for the ease of description: 
(a) An \textit{inline citation mention} is a paragraph from the main text of a paper that mentions another paper. For example, this paragraph from the RoBERTa paper~\citep{liu2019roberta}, ``\textit{... Unlike Devlin et al., (2019), ... we do not train with a reduced sequence length for the first 90\% of updates ...}'' mentions the BERT paper~\citep{devlin2019bert}.
(b) The \textit{source paper} is the paper the inline citation mention is sampled from. 
(c) A \textit{target paper} is a paper that is cited by the inline citation mention.

\Cref{fig:inline} provides an overview of our data collection methodology for \inlineq{} questions. We utilize the Semantic Scholar Open Research Corpus (S2ORC; \citealp{lo-etal-2020-s2orc}), a large corpus of academic papers obtained from publishers, archives, and the  Internet. We randomly sample inline citation mentions 
from the S2ORC\footnote{S2ORC provides detailed citation information for most inline citations (including the position of the citation in the paper text and the unique identifier of the target paper), enabling us to easily sample inline citation mentions and match them to S2ORC papers. We used the 2024-03-26 version.} and prompt 
GPT-4 to rewrite these citation mentions into literature search questions. 
These questions are filtered to remove those with a high word overlap with the title of the target papers, and are further manually examined to ensure high quality.

\paragraph{Sampling inline citation mentions.}
We limit the target papers to be only from the ACL Anthology, for the purpose of aligning with the expertise of the manual annotators, i.e., authors of this work. However, we do not limit where the source papers come from. Depending on the source papers, we call 
these questions \emph{ACL sourced} or \emph{non-ACL sourced}.

\begin{table}[t]
    \centering
    \small
    \begin{tabular}{lp{20em}}
        \toprule
        \multicolumn{2}{l}{\tf{Specificity}} \\
        \midrule
        0 & \emph{Broad}. There should exist no more than 20 papers that fit the question.\vspace{-5pt}\\
        & {\color{blue} Example: What are some parameter-efficient fine-tuning methods?}\vspace{5pt}\\ 
        1 & \emph{Specific}. There should exist no more than 5 papers that fit the question.\vspace{-5pt}\\
        & {\color{blue} Example: Which method involves training additional prompt tokens for every layer during the fine-tuning of language models?}\\
        \midrule
        \multicolumn{2}{l}{\tf{Quality}} \\
        \midrule
        0 & \emph{Discarded}. The question is factually wrong, unrealistic, overly broad/specific, or too easy. \\
        1 & \emph{Acceptable}. The question can be somewhat out of distribution of what researchers ask, or relatively easy due to high overlap with the title/abstract. \\
        2 & \emph{Good}. The question makes a challenging yet meaningful literature search question.\\
        \bottomrule
    \end{tabular}
    \caption{Annotation rubrics for the manual filtering (conducted by the authors of \ours{}).} %
    \label{tab:manual_filter}
\end{table}

\paragraph{Prompting GPT-4 to generate questions.}
Given a sampled citation mention,
we prompt GPT-4~\citep{openai2023gpt4} to generate a literature search question. In the prompt, we provide (1) the sampled paragraph (the inline citation mention) from the source paper and (2) the titles of the cited papers, and instruct GPT-4 to generate a literature search question based on the paragraph that would be answered by one or more of the papers cited in the paragraph.
We use in-context learning~\citep{brown2020language} and include two demonstrations. 
The prompt we use can be found in 
\Cref{tab:prompt_gen_question}. %

\paragraph{Word-overlap filtering.}
We notice that \inlineq{} questions generated in the last step can have very high word overlap with the target paper titles, which makes their retrieval trivial even for BM25 and suggests that the questions may not be of interest to researchers. %
We calculate the word overlap 
as the \textit{percentage of words in the generated question that are also included in the target paper titles}.
We filter out ACL sourced questions that have an overlap score higher than 0.3 and 
non-ACL sourced questions that have an overlap score higher than 0.1.
In this step, we filter out 5\% of the ACL sourced questions and 80\% of the non-ACL sourced questions.

\begin{table*}[t]
    \centering
    \small
    \setlength{\tabcolsep}{5pt}
    \begin{tabular}{lcccccccccc}
        \toprule
        &\multicolumn{4}{c}{\tf{Broad}} & \multicolumn{4}{c}{\tf{Specific}}  & \multirow{2}{*}{\tf{Total \#Q}} \\
        \cmidrule(lr){2-5} \cmidrule(lr){6-9} 
        & \#Q & Avg. Len & Overlap & Avg. \#P & \#Q & Avg. Len & Overlap & Avg. \#P\\
        \midrule
        \tf{\inlineQ{} Questions} & 120 & 20.6 & 0.33 & 1.21 & 231 & 22.1 & 0.34 & 1.07 & 351 \\
        \tf{\authorQ{} Questions} & 35 & 15.8 & 0.43 & 1.03 & 211 & 17.9 & 0.43 & 1.00 &246 \\
        \bottomrule
    \end{tabular}
    \caption{Statistics for \ours{}. Please refer to \Cref{table:detailed_stats} for more detailed statistics of each subset. ``\#Q'': number of questions. ``Overlap'': the fraction of words in the question that are also included in the titles and abstracts of the target papers. ``Avg. \#P'': average number of target papers.
    }
    \label{table:stats}
\end{table*}

\subsection{\Authorq{} Questions}
\label{sec:author}

Besides generating questions using existing inline citation mentions, we also collect questions directly from human annotators. 
As writing literature search questions  requires deep understanding of the research field and the target paper,
we invite researchers to write search queries that are answered by their own published papers. 
One additional benefit of this setup is that the correctness of the questions is better guaranteed.

We invited authors of ACL 2023 and ICLR 2024 papers to write one literature search question for each of their papers. We chose the two venues as they were among the latest natural language  and machine learning conferences at the time of the data collection, hence the papers represent the latest research development and are unlikely to have already been included in the pre-training data of LLMs and retrievers used in our evaluations.
We sent out invitations to $623$ ACL 2023 authors and $404$ ICLR 2024 authors, and received $175$ questions from ACL 2023 authors and $117$ questions from ICLR 2024 authors.

\subsection{Manual Filtering to Ensure High Quality}
\label{sec:filter}

Finally, the authors of this work manually examine every question from both the \inlineq{} and \authorq{} subsets and annotate these for \textit{specificity} and \textit{quality} (guidelines in \Cref{tab:manual_filter}). 
Questions that are too general (there are more than 20 papers from the corpus can fit the question) are assigned a quality score of 0 and are excluded. We include only questions with a quality score of 1 or 2 in the final dataset. 
We also rewrite questions if they have minor mistakes and can be fixed easily.
Each question is assigned to one author for annotation.

As the examples in \Cref{tab:manual_filter} show, 
questions of both specificity types can be realistic and valuable, but they exhibit distinct traits.
We use the specificity scores to distinguish \emph{broad} and \emph{specific} questions  in the evaluation.

For the \inlineq{} subset, 
we manually examined 382 ACL sourced questions and 450 non-ACL sourced questions.
26\% (98 instances) of the ACL questions and 
56\% (253 instances) of the non-ACL questions are kept. For the \authorq{} subset, since all questions are written by experts, we avoid rewriting them as much as possible. In the end, we kept 89\% (155) questions from ACL 2023 authors and 78\% (91) questions from ICLR 2024 authors.

\vspace{0.8em}
\subsection{Dataset Statistics}
Our final dataset contains 597 questions, with 351 in the \inlineq{} subset and 246 in the \authorq{} subset. Dataset statistics, including the number of questions, the average question length, and the average word overlap between the question and the target papers (titles and abstracts), are presented in  \Cref{table:stats}. 
We find that \authorq{}
questions are shorter and have a higher word overlap rate with the target papers (0.43 vs. 0.33 for \inlineq{} questions). 
This is expected: when writing questions for their own papers, 
authors tend to re-use terminology from their papers and focus on  the main findings which are usually included in the abstracts or titles. 
In contrast, \inlineq{} questions can be anchored to any span of the reference documents, irrespective of the main findings or the main focuses of the target papers.

\subsection{The Retrieval Corpus}
The \ours{} retrieval corpus $\mathcal{P}$ consists of ACL Anthology and ICLR papers extracted from 
S2ORC (see \Cref{app:corpus} for details).
We do not use the full S2ORC corpus for efficiency reasons.
In total, this yields $64,183$ papers ($59,383$ ACL Anthology papers and $4,807$ ICLR papers)\footnote{$7$ papers are common to both subsets.}. 
The average number of words for the documents in $\mathcal{P}$ is 134 / 6,041 (titles and abstracts / full texts).%

\section{Experiments}
\label{sec:systems}

\subsection{Experimental Setup} We compare the performance of different retrieval systems (enumerated below) on our \ours{} benchmark. Due to the limited context sizes of existing embedding models, we only use the paper titles and abstracts to embed the papers in our retrieval corpus $\mathcal{P}$ by default. 

For all systems we compare, we report the recall@K for both the broad and specific subsets of \ours{}. We report results for $K=5, 20$ for the specific subset and $K=20$ for the broad subset; these values (5 and 20) correspond to the guidelines followed by the authors while determining the specificity of a given question (see Table \ref{tab:manual_filter}).

\subsection{Baselines} We benchmark both retrieval models and LLM-based rerankers in this work. 

\paragraph{Retriever models.} We evaluate using the classic BM25 algorithm~\citep{bm25}, as well as several state-of-the-art dense retrieval (embedding) models,
including \gtr{}~\citep{ni-etal-2022-large},
\instructor{}~\citep{su-etal-2023-one},
\efive{}~\citep{Wang2022TextEB},
and \grit{}~\citep{muennighoff2024generative}.\footnote{We use the following corresponding checkpoints from \url{https://huggingface.co/}:
\gtrfull{}, \instructorfull{}, \efivefull{}, and \gritfull{}.}
More details are provided in \Cref{app:retriever}.

\begin{table*}[th]
    \centering
    \small
        \begin{tabular}{lcccccccc}
            \toprule
            & \multicolumn{3}{c}{\textbf{\inlineQ{}}} & \multicolumn{3}{c}{\textbf{\authorQ{}}} & \tf{Avg.} & \tf{Avg.}\\ 
            & \textbf{Broad} & \multicolumn{2}{c}{\tf{Specific}} & \textbf{Broad}      & \multicolumn{2}{c}{\tf{Specific}}  & \tf{Broad} & \tf{Specific}    \\
            \cmidrule(lr){2-2}  \cmidrule(lr){3-4} \cmidrule(lr){5-5} \cmidrule(lr){6-7} \cmidrule(lr){8-8} \cmidrule(lr){9-9}
            & R@20 & R@5 & R@20 & R@20 & R@5 & R@20 & R@20 & R@5 \\
            \midrule
            BM25  & 37.4 & 38.5 & 55.8 & 48.6 & 62.6 & 73.5 & 39.9 & 50.0 \\
            \gtrfull{}   & 45.7 & 38.5 & 51.5 & 37.1 & 40.8 & 55.9 & 43.8 & 39.6 \\
            \instructorfull{} & 56.3 & 48.9 & 60.0 & 57.1 & 55.9 & 70.1 & 56.5 & 52.3 \\
            \efivefull{} & 55.8 & 50.4 & 63.9 & 54.3 & 62.6 & 75.8 & 55.4 & 56.2 \\
            \gritfull{} & \tf{69.7} & \tf{67.7} & \tf{77.9} & \tf{74.3} & \tf{82.5} & \tf{89.1} & \tf{70.8} & \tf{74.8} \\
            \midrule
            GPT-4o reranking (w/ BM25) & 54.9 & 60.0 & 67.5 & \tf{77.1} & 76.8 & 82.9 & 59.9 & 68.0 \\
            GPT-4o one-hop (w/ BM25) & 62.0 & 64.1 & 71.6 & 74.3 & 73.5 & 77.7 & 64.8 & 68.6 \\
            GPT-4o reranking (w/ \grit{}) & \tf{74.7} & \tf{73.2} & \tf{79.9} & \tf{77.1} & \tf{85.8} & \tf{92.4} & \tf{75.3} & \tf{79.2} \\
            GPT-4o one-hop (w/ \grit{}) & 72.9 & 70.3 & 78.4 & 74.3 & 84.4 & 87.2 & 73.2 & 77.0 \\
            \bottomrule
       \end{tabular}
    \caption{
        Main experimental results of \ours{}.
        Here we only use the titles and abstracts of papers for retrieval and reranking.
        We report recall@20 (R@20) for broad questions and
        recall@5 and @20 (R@5, R@20) for specific questions.
        ``Broad'' and ``specific'' correspond to the annotations during our manual filtering stage (defined in \Cref{tab:manual_filter}).
    }
    \label{table:main_result}
\end{table*}

\begin{figure*}[ht]
    \centering
    \includegraphics[width=0.98\linewidth]{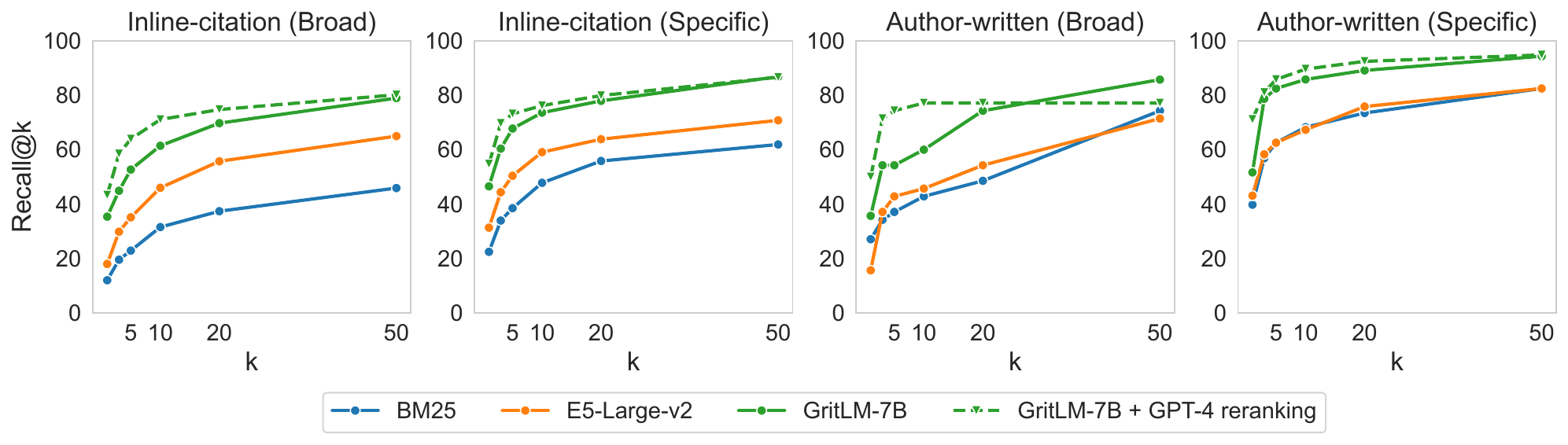}
    \caption{We demonstrate detailed retrieval results using BM25, E5 and \grit{} up to $k=50$. Additionally, we show the effect of applying GPT-4o reranking over \grit{} retrieval results.
    } 
    \label{fig:query_set_recalls}
\end{figure*}

\paragraph{LLM-based reranking.}
In addition to vanilla retrieval, we also use strong LLMs (GPT-4o\footnote{We use gpt-4o-2024-05-13 in our experiments.} in our case) to rerank the top retrieved results from the above retrievers. We use two strategies:

 \textbf{Vanilla reranking.} We include the top-$n$ retrieved papers (titles and abstracts) in the context and prompt GPT-4o to rerank these based on the question (see our prompt in \Cref{tab:prompt_rerank}). This is similar to prior works \citep{sun-etal-2023-chatgpt,ma2023zero}. We use $n=100$, resulting in an average context length of 13,844 words. 

\textbf{One-hop reranking.} Inspired by \citet{tang2023referral}, we leverage the fact that for some questions, there may exist lexically similar inline citation mentions in the retrieval corpus. Due to our data collection pipeline, this is particularly true for the ACL sourced inline citation questions. %
We posit that the retrieval models  will be able to retrieve these source papers (that cite the target papers) based on the questions. 

We extract the top $m$ retrieved papers and construct a new candidate list by adding papers cited by each of these seed retrieved papers. We concatenate papers in the following order, skipping duplicates: [rank-1 paper $p_1$, papers cited by $p_1$, rank-2 paper $p_2$, papers cited by $p_2$, ..., $p_m$, papers cited by $p_m$].
To avoid very long contexts, we truncate this list after the first $n$ papers and use the same prompt for GPT-4o based reranking as above. In our experiments, we use $m=50$ and $n=200$ resulting in average length of 27,544 words.

\subsection{Results}
\label{sec:main_result}

We outline the performance of the above systems on \ours{} in \Cref{table:main_result}. First, we observe that all instruction-finetuned embedding models, e.g.  \instructor{}, \efive{}, and \grit{}, substantially outperform BM25 on our benchmark. In fact, they also perform better than the \gtr{} model. Overall, we found that \gritfull{} achieves the best performance ($70.8$ recall@20 on broad questions and $74.8$ recall@5 on specific questions), leaving a large gap compared to other baselines.

\paragraph{Impact of reranking.}
We also report the performance improvement brought by the reranking methods on the weakest (BM25) and strongest (\grit{}) retrievers  in \Cref{table:main_result}. We observe that both vanilla and one-hop reranking improve over the base retrieval performance. For example, on the specific subset of inline questions, the vanilla GPT-4o reranking improves the recall@5 of BM25 and \grit{} by $21.5\%$
and $5.5\%$ respectively. 
Interestingly, the improvements from one-hop reranking are generally lower than vanilla reranking across all subsets, when using \grit{} as the base retriever; this shows that our benchmark  cannot be easily ``gamed'' by mimicking the data collection pipeline or exploiting similar citation mentions to the question from other papers.  %

\paragraph{Impact of question specificity.} \Cref{table:main_result} and \Cref{fig:query_set_recalls} show that retrieval systems generally report higher recall performance on the specific subset. This is expected: there exists a smaller number of ``competing'' papers, i.e. those that also satisfy the search question, for the specific subset in the retrieval corpus. 
Note that our human annotation tagged questions with approximately 5 relevant papers  as specific and 20 relevant papers as broad (see \Cref{tab:manual_filter} for details). 
We keep both subsets in our dataset as this stratified reporting presents a more nuanced view of retriever capabilities. %

\paragraph{\inlineQ{} vs. \authorq{} questions.} 
We observe very different performance trends for the two subsets  (\Cref{table:main_result} and \Cref{fig:query_set_recalls}). 
In particular, inline-citation questions are harder than author-written questions for all retriever systems, on both broad and specific questions.

We attribute this difference to the higher semantic or lexical overlap of \authorq{} questions with the paper titles and abstracts (see \Cref{table:stats} for statistics). 
This reflects expected tendency of paper authors to formulate the questions around the main contributions from the abstracts and re-use terminologies. Such annotator biases have been widely discussed in prior data collection efforts as well, particularly when humans write content from scratch~\citep{gururangan-etal-2018-annotation}.

\begin{table}[t]
    \centering
    \small
    \resizebox{0.98\linewidth}{!}{
        \begin{tabular}{lcccc}
            \toprule
            & \multicolumn{2}{c}{\textbf{Inline (specific)}} & \multicolumn{2}{c}{\textbf{Author (specific)}} \\
            \cmidrule(lr){2-3} \cmidrule(lr){4-5}
             & \textbf{Qual=1} & \textbf{Qual=2} & \textbf{Qual=1} & \textbf{Qual=2} \\
            & R@5 & R@5 & R@5 & R@5 \\
            \midrule
            BM25  & 36.4 & 30.6 & 62.2 & 55.0 \\
            \gtrfull{}   & 42.0 & 31.4 & 40.7 & 36.9 \\
            \instructorfull{} & 55.1 & 39.5 & 58.5 & 48.6 \\
            \efivefull{} & 48.4 & 42.6 & 61.5 & 57.7 \\
            \gritfull{} & 67.3 & 58.7 & 80.0 & 76.6 \\
            \bottomrule
       \end{tabular}
    }
    \caption{
       Comparison of retrieval performance on different quality (qual) questions. Generally, retrievers report lower performance on the Qual=2 questions, i.e. those deemed more challenging in our manual annotation. 
    }
    \vspace{-5pt}
    \label{table:result_question_quality}
\end{table}

\paragraph{Impact of question quality.}
\label{sec:quality}

\Cref{table:result_question_quality} compares how retrieval models perform on different quality subsets. Recall that we manually annotated the quality of all questions (\Cref{tab:manual_filter}). We observe that questions with a quality score of $2$, i.e. determined to be more realistic and difficult by manual annotators, are consistently more challenging for retrieval models. This demonstrates the high annotation quality of our manual inspection step. The presence of these different quality questions in our dataset leads to higher diversity and better coverage over the varied information seeking needs of users.

\section{Analysis}
\label{sec:analysis}

\begin{table}[t]
    \centering
    \small
        \begin{tabular}{rcccc}
            \toprule
            & \multicolumn{2}{c}{\textbf{Inline}} & \multicolumn{2}{c}{\textbf{Author}} \\
            \cmidrule(lr){2-3} \cmidrule(lr){4-5}
            & \textbf{Broad} & \textbf{Spec} & \textbf{Broad} & \textbf{Spec} \\
            & R@20 & R@5 & R@20 & R@5 \\
            \midrule
            BM25 & \tf{37.4} & \tf{38.5} &  48.6 & 62.6 \\
            ~~~w/ full  & 18.6 & 23.8  & \tf{65.7} & \tf{71.6} \\ \midrule
            \gtrfull{}   & \tf{45.7} & 38.5 &  37.1 & \tf{40.8} \\
            ~~~w/ full  & 43.9 & \tf{39.4} &  \tf{45.7} & 39.8 \\ \midrule
            \instructorfull{} & \tf{56.3} & 48.9  & \tf{57.1} & 55.9 \\
            ~~~w/ full  & 53.0 & \tf{50.9}  & \tf{57.1} & \tf{56.9} \\ \midrule
            \efivefull{} & 55.8 & \tf{50.4}  & 54.3 & \tf{62.6} \\
            ~~~w/ full  & \tf{56.9} & 48.7 &  \tf{60.0} & 62.1 \\ \midrule
            \gritfull{} & {69.7} & \tf{67.7}  & \tf{74.3} & \tf{82.5} \\
            ~~~w/ full  & \tf{70.8} & 63.4  & 65.7 & 73.0 \\
            \bottomrule
       \end{tabular}
    \caption{
        Retrieval results of using only titles and abstracts vs. using titles, abstracts, and full text (w/ full). We do not observe consistent improvements from including the full text for existing retrieval models. %
    }
    \vspace{-5pt}
    \label{table:fulltext_vs_abstract}
\end{table}

\subsection{Does Including More Paper Content Improve Retrieval Performance?}
\label{sec:full_vs_abstract}

In the previous section, we only used the titles and abstracts (on average 134 words) to encode the papers in the retrieval corpus. Here, we evaluate whether encoding more paper content can improve retrieval performance. For all retriever models compared, we create embeddings using the full paper text (on average 6,041 words) up to their allowed context lengths.\footnote{The maximum context lengths for  \gtrfull{}, \instructorfull{}, \efivefull{} and \gritfull{} are 512, 512, 512 and 2048 tokens respectively.} We compare this setting against our default setting (only titles and abstracts).

Our results are outlined  in \Cref{table:fulltext_vs_abstract}. Surprisingly, we find that the addition of more paper text does not improve performance on \ours{} consistently. In fact, we only observe substantial improvement on the \authorq{} broad questions for BM25 and some embedding models. In other cases, more text more often hinders instead of improving performance. 
Note that the maximum context length of the tested models is 2,048 (\grit{}) and the average length of their training data is even shorter---for example, the commonly used MS-MARCO~\citep{nguyen2017ms} and NaturalQuestions~\citep{lee-etal-2019-latent} have an average document length of 56 and 79. This is  significantly shorter than the full text of papers from our retrieval corpus averaging 6,041 words in length, potentially leading to the
unsatisfying performance when using full texts with embedding models.

\begin{table}[t]
    \centering
    \small
    \resizebox{0.98\linewidth}{!}{
        \begin{tabular}{lcccc}
            \toprule
            & \multicolumn{2}{c}{\textbf{ACL}} & \multicolumn{2}{c}{\textbf{Non-ACL}} \\
            \cmidrule(lr){2-3} \cmidrule(lr){4-5}
            & \textbf{Broad} & \textbf{Specific} & \textbf{Broad} & \textbf{Specific} \\
            & R@20 & R@5 & R@20 & R@5 \\
            \midrule
            BM25  & 38.8 & 39.4 & 36.9 & 38.2 \\
            \gtrfull{}   & 37.2 & 39.4 & 48.9 & 38.2 \\
            \instructorfull{} & 48.6 & 43.9 & 59.1 & 50.9 \\
            \efivefull{} & 46.6 & 46.2 & 59.1 & 52.1 \\
            \gritfull{} & 72.4 & 65.9 & 68.8 & 68.5 \\
            \midrule 
            \tf{With BM25} \\
            ~~~Reranking & 51.1 & 59.8 & 56.2 & 60.0 \\ 
            ~~~One-hop & \tf{65.2} & \tf{71.2} & \tf{60.8} & \tf{61.2} \\
            \tf{With \grit{}} \\
            ~~~Reranking  & 80.3 & \tf{72.7} & \tf{72.7} & \tf{73.3} \\
            ~~~One-hop  & \tf{81.3} & 67.4 & 69.9 & 71.5 \\
            \bottomrule
       \end{tabular}
    }
    \caption{Comparison of retrieval performance on the ACL vs. non-ACL sourced \inlineq{} questions. Results show that the performance improvement from one-hop reranking over BM25 is subtantially higher for ACL sourced questions.}
    \label{table:acl_vs_nonacl}
\end{table}

\subsection{Does the Source of Inline Citation Questions Matter?}
\label{sec:inlinequestion_acl}

Next, we study how the different sources of \inlineq{} questions
affect the model performance. \Cref{table:acl_vs_nonacl} outlines the performance of retrieval models on ACL sourced vs. non-ACL sourced \inlineq{} questions. 
Our results show that the two different sets report similar trends and model rankings for different retrieval models, particularly on the specific subset of questions. Interestingly, we find that the performance improvement from one-hop reranking is very different for the ACL and non-ACL questions. 

For BM25, we observe that one-hop reranking is significantly better than the vanilla reranking on ACL sourced questions (+11.4\% recall@5 on specific); but the gap is much smaller on non-ACL sourced questions (+1.2\% recall@5 on specific). 
We posit that this is because BM25 can better exploit the data annotation pipeline on the ACL sourced questions. It can likely first identify the source ACL paper where the citation mention comes from, and then find the target paper via one-hop reranking. Including the non-ACL questions to \ours{} prevents systems from exploiting such ``shortcuts'' as non-ACL source papers are not part of the retrieval corpus. 

For \grit{}, we do not observe similarly large performance gains when using one-hop reranking. We hypothesize that this is because when using \grit{}, the initial top retrieval results already include the target papers and the one-hop strategy does not bring further improvement.

\begin{table}[t]
    \centering
    \small
        \begin{tabular}{lcc}
            \toprule
            & \multicolumn{1}{c}{\textbf{Inline (specific)}} & \multicolumn{1}{c}{\textbf{Author (specific)}} \\
            & R@5 & R@5 \\
            \midrule
            BM25  & 38.5 & 62.6 \\
            \gritfull{} & 67.7 & 82.5 \\ \midrule
            Google Search& 23.1 & 62.5\\
            Google Scholar & 20.5 & 17.5 \\
            Elicit & 23.1 & 17.5 \\
            \bottomrule
       \end{tabular}
    \caption{Recall@5 for commercial search engines on a random subset of 80 specific questions. Search engines generally report poor performance. Note that the comparison is not apples-to-apples as search engines use a much larger retrieval corpus.}
    \label{table:search_engines}
\end{table}

\subsection{Performance of Search Engines}
\label{sec:search_engines}

\begin{table*}[th]
    \centering
    \small
        \begin{tabular}{lccccccc}
            \toprule
            & \tf{MSMARCO} & \tf{SCIDOCS} & \tf{NQ} & \tf{ArXiv} & \tf{\ours{}} (broad) & \tf{\ours{}} (specific) \\
            \midrule
            \gtrfull{}   & 42.7 & 15.5 & 55.1 & 17.5 & 23.3 & 30.4 \\
            \instructorfull{} & 41.6 & 17.4 & 57.2 & 19.8 & 32.8 & 41.2 \\
            \efivefull{} & 43.5 & 20.5 & 63.4 & 27.0 & 27.1 & 45.3 \\
            \gritfull{} & 42.0 & 24.4 & 70.3 & 34.3 & 44.1 & 60.3 \\
            \bottomrule
       \end{tabular}
    \caption{
        Comparison between \ours{} and existing retrieval benchmarks.
        All reported numbers are  nDCG@10 for a direct comparison.
    }
    \vspace{-5pt}
    \label{table:vs_existing_benchs}
\end{table*}

In practice, researchers use search engines like Google Search, Google Scholar, or Elicit\footnote{\url{https://elicit.com/}} to search for relevant papers for their scientific queries. We conduct a human study to understand how these search engines perform on \ours{}: We randomly sample 80 questions (all specific; 40 \inlineq{} and 40 \authorq{}) from our dataset. We manually input\footnote{We use  incognito mode browser sessions, but do not specially prevent IP-address-based location tracking.} these questions into the above search engines and report recall@5.\footnote{For Google Search, we consider the top-5 academic papers in the search results and ignore other webpage results. We only consider the academic papers on the first page of search results, even if this number is lower than 5.} We note that this is not an apples-to-apples comparison against the retrieval models in earlier sections due to the discrepancy in the retrieval corpus.

\Cref{table:search_engines} outlines the results of our human study. It shows that all three search engines deliver similarly low recalls on \inlineq{} questions. On the \authorq{} questions, Google Search performs much better than the other two. Although not directly comparable, this performance is generally worse than the embedding models, demonstrating the potential of these strong dense retrieval models for citation recommendation  applications.

\subsection{Comparing Other Retrieval Benchmarks}
\label{sec:comp_existing_benchmarks}

We compare model performance on \ours{}
to several popular retrieval benchmarks included in BEIR~\citep{thakur2021beir} and MTEB~\citep{muennighoff2022mteb}---namely MSMARCO~\citep{nguyen2017ms}, SCIDOCS~\citep{cohan-etal-2020-specter}, and NQ~\citep{lee-etal-2019-latent}.\footnote{We use the results from the MTEB benchmark website: \url{https://huggingface.co/spaces/mteb/leaderboard}.} 
We also compare to ArXiv~\citep{gu2022local}, a previous citation recommendation benchmark directly using inline citations as queries. 
\Cref{table:vs_existing_benchs} shows that 
\ours{} generally agrees with existing retrieval benchmarks.
However, 
\ours{} can differentiate retriever models better: for example, the gap between \grit{} and \efive{} on \ours{} (specific) is 15 points (nDCG@10), while they perform almost the same on MSMARCO. 
\ours{} provides an informative testbed that can effectively reflect the recent (and future) advancement in embedding models.

\section{Related Work}
\label{sec:relatedwork}

\paragraph{Citation recommendation.} 
The community has proposed a number of citation recommendation datasets~\cite{farber2020citation},
including global citation recommendation datasets (directly using a paper as the query and papers it cites as target papers; \citealp{cohan-etal-2020-specter,bhagavatula-etal-2018-content}), and 
local citation recommendation datasets (using inline citation mentions as queries; ~\citealp{he2010context,medic-snajder-2020-improved,Jeong2020,gu2022local}).
There are also language models and retrieval models specifically trained for scientific document understanding and retrieval tasks, such as SciBERT~\cite{beltagy-etal-2019-scibert} and SPECTER~\cite{cohan-etal-2020-specter}.
Compared to existing citation recommendation datasets,
\ours{} is comprised of manually annotated, natural language literature search questions,
providing a more realistic and challenging evaluation for citation recommendation systems.

\paragraph{Retrieval benchmarks.}
There have been numerous datasets evaluating 
retrieval systems from
Wikipedia~\citep{kwiatkowski2019natural,lee-etal-2019-latent},
web queries~\citep{nguyen2017ms}, 
biomedical questions~\citep{voorhees2000building_trec}, and more.
Recently, there have been several benchmarks
combining multiple datasets 
and evaluating retrieval or embedding models 
across different domains and different use cases, such as KILT~\citep{petroni-etal-2021-kilt}, BEIR~\citep{thakur2021beir}, and MTEB~\citep{muennighoff2022mteb}.
\ours{} offers a unique perspective %
by exploring the novel literature search question type, 
effectively complementing the existing benchmarks.
Contemporary with our work, CiteME~\citep{press2024citemelanguagemodelsaccurately} introduces a benchmark for identifying references based on claims made in a paper's inline text (as opposed to research questions in our case). Their focus differs from ours in that it is aimed more at LLM-based agents rather than retrieval systems.

\paragraph{Retrieval systems.} 
Traditional retrieval systems rely on bag-of-word algorithms such as TF-IDF and BM25. 
Dense retrieval (embedding) models have gained more popularity due to their abilities to do semantic search without relying on exact keyword matches~\cite{pennington-etal-2014-glove,reimers2019sentence}.
State-of-the-art dense models 
are mostly adopted by
fine-tuning pre-trained language models~\citep{devlin2019bert,touvron2023llama}
with a contrastive learning objective on either supervised or unsupervised data~\citep{karpukhin-etal-2020-dense,gao-etal-2021-simcse,izacard2022unsupervised,ni-etal-2022-large, Khattab2020ColBERTEA}.
Recent development introduces ``instructions'' when encoding  queries and  documents, which significantly improves the versatility of embeddings across  tasks~\citep{su-etal-2023-one, Wang2022TextEB, wu2022grit, lee2024nvembed, behnamghader2024llm2vec}.

\section{Conclusion}
In this paper, we propose \ours{},
a new retrieval benchmark comprising 597 manually-curated literature search questions.
\ours{} includes an \inlineq{} question set and an \authorq{} question set, both undergoing manual inspection from the authors of \ours{}.
We conduct extensive experiments with BM25, state-of-the-art embedding models, and LLM reranking. %
Our experiments demonstrate the superior performance of state-of-the-art instruction-finetuned embedding models, 
with additional improvement via GPT-4o-based reranking.
We also verify that commercial search engines like Google struggle with \ours{} questions.
The comparison with existing retrieval benchmarks shows that 
\ours{} better differentiates the performance of retrieval systems.

\section*{Limitations}
Even though we manually examined the dataset,
there still exist questions that are either slightly out of distribution compared to what researchers would ask, 
or too easy due to high overlap with the target papers.
The \authorq{} questions are easier than we expected,
as writing challenging literature search questions is non-trivial even for experienced researchers. 
Even though we experimented with several state-of-the-art systems, it was not an exhausted evaluation and we left out more sophisticated retrieval or reranking systems. 
This research primarily focuses on only English questions and research papers.

\section*{Ethics Statement}

The research artifact of this paper, \ours{}, is manually inspected and has been ensured to have no unsafe or inappropriate  content. 
However, the process to generate the dataset may introduce certain biases: for example, the \inlineq{} questions contain more target papers that have high citations due to the sampling; the \authorq{} questions only cover ACL 2023 and ICLR 2024 papers.

\section*{Acknowledgements}
We want to acknowledge Dan Friedman, Howard Yen, Jiayi Geng, Lucy He, and other members of the Princeton NLP group for their useful feedback and discussion.
We also acknowledge all the ACL 2023 and ICLR 2024 authors that contributed questions to \ours{} (listed in \Cref{app:author_ack}). 
Tianyu Gao is supported by an IBM PhD Fellowship.
This work is gratefully supported by an NSF CAREER award (IIS-2239290), and Microsoft Azure credits through the ``Accelerate Foundation Models Academic Research'' Initiative.

\bibliography{custom}

\clearpage
\appendix

\onecolumn

\section{Annotator Acknowledgments}
\label{app:author_ack}

We would like to thank Marah I Abdin, Jaewoo Ahn, Kabir Ahuja, Xi Ai, Satoshi Akasaki, Anastasios N Angelopoulos, Jinheon Baek, Eslam Mohamed Bakr, Pablo Barceló, Claudio Battiloro, Jonas Belouadi, Abhik Bhattacharjee, Valeriia Bolotova, Pengshan Cai, Nitay Calderon, Qingqing Cao, Defu Cao, Souradip Chakraborty, Jun Shern Chan, Sachin Chanchani, Yulong Chen, Yiming Chen, Xinyuan Chen, Nuo Chen, Hanjie Chen, Xiudi Chen, Zeming Chen, An-Chieh Cheng, Xize Cheng, Cheng-Han Chiang, Josef Dai, David Dale, Yue Deng, Yifan Deng, Shizhe Diao, Bosheng Ding, Xuan Long Do, Yilun Du, Yupei Du, Salijona Dyrmishi, Dante Everaert, Zhenghan Fang, Bahare Fatemi, Jiazhan Feng, Shangbin Feng, Patrick Fernandes, Javier Ferrando, Christopher Fifty, Sarah E Finch, Matthew Finlayson, Lea Frermann, Mikhail Galkin, Songyang Gao, Ziteng Gao, Silin Gao, Sara Ghazanfari, Nathan Godey, Navita Goyal, Xinran Gu, Yuxian Gu, Yu Gu, Anchun Gui, Jiacheng Guo, Ashim Gupta, Paul Hagemann, Tianxing He, Zhengfu He, Juncai He, Leonhard Hennig, Konstantin Hess, Jennifer Hu, Xiaoyang Hu, Zhilei Hu, Weidong Huang, Yichong Huang, Ayyoob Imani, Qi Jia, Yifan Jiang, Hanwen Jiang, Yiding Jiang, Yang Jin, Youngjin Jin, Zhijing Jin, Emmeran Johnson, Josef Jon, David Jurgens, Ehsan Kamalloo, Junmo Kang, Jian Kang, Mikhail Khodak, Hyunjae Kim, Soroush Abbasi Koohpayegani, Suhas Kotha, Jeongyeol Kwon, Sunjae Kwon, Philippe Laban, Zhibin Lan, Nayoung Lee, Deokjae Lee, Celine Lee, Heejun Lee, Jie Lei, Wenhao Li, Yafu Li, Yufei Li, Yanzeng Li, Yanzhou Li, Ziqiang Li, Zhaoyi Li, Ziheng Li, Xiaonan Li, Yinghao Li, Yu Li, Chengrui Li, Yingjie Li, Yunlong Liang, Baohao Liao, Kezhou Lin, Licong Lin, Enrico Liscio, Xiangyan Liu, Chenzhengyi Liu, Yixin Liu, Xingbin Liu, Haolin Liu, Xiao Liu, Yajiao Liu, Meng Liu, Tianyang Liu, Wei Liu, Qingyu Lu, Pan Lu, Junyu Lu, Zhengyi Luo, Yang Luo, Ang Lv, Junhyung Lyle, Jiajun Ma, Kaixin Ma, Ziqiao Ma, Mounica Maddela, Chaitanya Malaviya, Zhiyu Mei, Ethan Mendes, Fatemehsadat Mireshghallah, Niloofar Mireshghallah, Mircea Mironenco, Takeru Miyato, Fengran Mo, Xinyi Mou, Niklas Muennighoff, Cheolwon Na, Piotr Nawrot, Mang Ning, Longshen Ou, Siqi Ouyang, Lorenzo Pacchiardi, Ziqi Pang, Sara Papi, Letitia Parcalabescu, Tanmay Parekh, Aleksandar Petrov, Lucía Pitarch, Moritz Plenz, Manish Prajapat, Joan Puigcerver, Valentina Pyatkin, Shuofei Qiao, Yujia Qin, Chengwei Qin, Sigal Raab, Hossein A Rahmani, Siyu Ren, Yubing Ren, Ruiyang Ren, Yangjun Ruan, Michael J Ryan, Shoumik Saha, Vageesh Saxena, Michael Saxon, Alexander Scarlatos, Agam Shah, Erfan Shayegani, Behzad Shayegh, Xiangqing Shen, Sheng Shen, Ruizhe Shi, Zhengliang Shi, Kensen Shi, Ziyi Shou, Prasann Singhal, Jasivan Alex Sivakumar, Junru Song, Chunjin Song, Nikita Srivatsan, Michal Štefánik, Hao Sun, Mingjie Sun, Weiwei Sun, Zhiqing Sun, Xiaohang Tang, Liyan Tang, Eshaan Tanwar, Jiayan Teng, Davide Testa, Changyao Tian, Yufei Tian, Eric Todd, Benjamin Towle, Austin Tripp, Yi Tu, Rheeya Uppaal, Lazar Valkov, Neeraj Varshney, Artem Vazhentsev, Yiming Wang, Qifan Wang, Zhaoyang Wang, Lirui Wang, Zhicheng Wang, Weiqi Wang, Jiaan Wang, Boshi Wang, Haiming Wang, Huimin Wang, Yun-Cheng Wang, Runzhe Wang, Yu Wang, Yidong Wang, Licheng Wen, Te-Lin Wu, Yu-Yu Wu, Qianhui Wu, Dongming Wu, Tong Wu, Zijun Wu, Mengzhou Xia, Jian Xie, Yiming Xie, Weiwen Xu, Yi Xu, Xilie Xu, Derek Xu, Shohei Yamasaki, Hao Yan, Chenghao Yang, Xianjun Yang, Sen Yang, Bingsheng Yao, Qinyuan Ye, Fan Yin, Haneul Yoo, Kiyoon Yoo, Xinyan Velocity Yu, Jianfei Yu, Qiying Yu, Mo Yu, Zichun Yu, Yue Yu, Youliang Yuan, Zihao Yue, Xiang Yue, Yuanwen Yue, Daoguang Zan, Zhiyuan Zeng, Guangtao Zeng, Yuheng Zha, Runzhe Zhan, Jiaxu Zhang, Zhexin Zhang, Chen Zhang, Xinlu Zhang, Yabo Zhang, Renrui Zhang, Kechi Zhang, Ruoyu Zhang, Feng Zhang, Siyan Zhao, Junhao Zheng, Wenjie Zheng, Ming Zhong, Yan Zhou, Pei Zhou, Yangqiaoyu Zhou, Aojun Zhou, Xuekai Zhu, Luyao Zhu, Yanqiao Zhu, Dele Zhu, Andrew Zhu, Wenjie Zhuo and Caleb Ziems for contributing author-written questions about their ACL 2023 and/or ICLR 2024 papers.

\section{Annotation Details}

We provide instructions regarding manually inspecting questions in Table~\ref{tab:manual_filter}. 
We sent out emails and Google Forms to recruit ACL 2023 and ICLR 2024 authors for \authorq{} questions, and 
the templates can be found in Table~\ref{tab:author_email} and Table~\ref{tab:author_form} respectively.

\section{Retrieval Corpus}
\label{app:corpus}

The \ours{} retrieval corpus $\mathcal{P}$ consists of ACL Anthology and ICLR papers extracted from 
S2ORC. Here we describe how we identify those papers in S2ORC:
We isolate ACL anthology papers from S2ORC by identifying entries whose metadata includes an ACL Anthology ID. We identify ICLR papers utilizing a combination of the venue-based queries to Semantic Scholar's Academic Graph API
and by title-matching using titles of accepted papers scraped from the official ICLR website. 

\section{Retriever details}
\label{app:retriever}

We list the full HuggingFace checkpoint paths corresponding to the dense retrievers we use in \Cref{tab:retriever_paths}. 
We use the following instructions for the instruction-finetuned embedding models: ``\texttt{Represent the research question for retrieving relevant research paper abstracts:}'' for encoding queries;  ``\texttt{Represent the title and abstract of the research paper for retrieval:}'' for encoding papers when using \instructorfull{} for retrieval using paper titles and abstracts; when performing retrieval using paper titles and abstacts with \gritfull{}, we use the instruction ``\texttt{Given a research query, retrieve the title and abstract of the relevant research paper}''.

\begin{table}[ht]
    \centering
    \small
    \begin{tabular}{cc}
        \toprule
        \tf{Retriever} & \tf{HuggingFace Checkpoint} \\
        \midrule
        \gtrfull{}   & \texttt{sentence-transformers/gtr-t5-large}   \\
        \instructorfull{} & \texttt{hkunlp/instructor-xl} \\
        \efivefull{} & \texttt{intfloat/e5-large-v2} \\
        \gritfull{} & \texttt{GritLM/GritLM-7B} \\
        \bottomrule
    \end{tabular}
    \caption{HuggingFace checkpoints we use for each dense retriever.}
    \label{tab:retriever_paths}
\end{table}

\section{Prompts and Additional Statistics} %
\label{app}

\Cref{tab:prompt_gen_question} shows the prompt we use for generating \inlineq{} questions via GPT-4.
\Cref{tab:prompt_rerank} shows the reranking prompt for GPT-4o.
\Cref{table:detailed_stats} shows a more detailed statistics about \ours{}.

\begin{table*}[h]
    \centering
    \small
    \begin{tabular}{>{\raggedright\arraybackslash\tt}p{0.98\textwidth}<{}}
        \toprule
            \vspace{-1em}
            Hi \{\{annotator name\}\}, \\ \\

            We hope this email finds you well! \\ \\
            
            First, congrats on your paper’s acceptance to \{\{conference name\}\}! We are [REDACTED] from [REDACTED] who are working on constructing a new challenging retrieval benchmark where the task is to retrieve relevant research papers given a research query. Would you be willing to dedicate 2 minutes to write a literature-search question about your \{\{conference name\}\} paper?  Here's the link to the google form: \{\{link\}\}. \\ \\
            
            Your contribution will help us build better, more challenging evaluations for large language models. We will make sure to list you as a contributor to our benchmark (unless you prefer otherwise). Thank you! \\ \\
            
            Best, \\
            \{\{author 1\}\} \\
            \{\{author 2\}\} \\
        \bottomrule
    \end{tabular}
    \caption{
        Email template sent out to ICLR 2024 and ACL 2023 authors for collecting author-written questions.
    }
    \label{tab:author_email}
\end{table*}

\begin{table*}[h]
    \centering
    \small
    \begin{tabular}{>{\raggedright\arraybackslash\tt}p{0.98\textwidth}<{}}
        \toprule
            \vspace{-1em}

Dear \{\{author name\}\}, \\ \\

Thanks for contributing to our literature-search question collection effort!  \\ \\

We are trying to create a new challenging retrieval benchmark where the task is to retrieve relevant research papers given a research query. For example, the research query "Which paper first found that large language models can do in-context learning?" should retrieve the GPT-3 paper. We could use your help for creating such a query that should be answered by your own paper! \\ \\

Could you provide one high-quality, literature-search-type query about your paper that you think would be challenging even for state-of-the-art retrieval systems? Please make sure to follow these guidelines when writing your query. Your query must be \\ \\

(1) Realistic: It should be plausible that a researcher working in a related field may ask this exact question. Do not ask a question that puts too many unrealistic constraints (such as "What work that uses the NaturalQuestions dataset trains for 15 epochs and uses a learning rate of 3e-5?"). \\ \\

(2) Specific: Your query should be answered by/correspond to one particular (representative) paper or a small number of papers. Do not ask an overly broad question like "What are some multimodal models?" Instead, you can ask "Which multimodal model was the first to use interleaved image and text data?" \\ \\

(3) Challenging: Our goal is to collect a set of queries that are extremely challenging even for SOTA retrievers today. Make sure not to submit a query that can easily be answered via keyword matching or a Google search like "Which paper proposed masked language modeling?" (BERT). Instead, you can use the following formats to make the question more challenging: \\
    (a) Asking a detailed question that the abstract does not cover: "Which text pre-training paper first used the data mixture of wikipedia and bookscorpus?" (BERT) \\
    (b) Rephrasing (reducing word overlap to the paper): "Is there such a machine learning dataset, where for some questions, there is no correct answer and model should abstain?" (SQuAD v2) \\
    (c) Asking about the main technique innovation: "Is there any paper that combines distillation and structured pruning for language models?" (CoFi pruning) \\ \\

Thank you for your contribution! We will make sure that: \\
    - Your name will be credited in our paper unless you choose otherwise. \\
    - We will make the dataset available for open-source development. \\

\bottomrule
    \end{tabular}
    \caption{
        Instructions provided in the Google Forms  sent to ICLR 2024 and ACL 2023 authors for collecting author-written questions.
    }
    \label{tab:author_form}
\end{table*}

\begin{table*}[ht]
    \centering
    \small
    \begin{tabular}{>{\raggedright\arraybackslash\tt}p{0.98\textwidth}<{}}
        \toprule
            \vspace{-1em}
            I will provide you with a excerpt from an scientific article that cites various papers in the positions \{cite\_001\}, \{cite\_002\}, etc. I would like you to write *general* questions about the academic literature using this paragraph, which will be answered with the corresponding citations. Here are the constraints you must follow: \\ 
            (1) The questions should make sense and be interesting without reading the paragraph. The questions should be general, and you should only use the paragraph to find the ground-truth answer. \\
            (2) Try to ask questions which cover as many citations as possible and are as detailed as possible. Pack as much information into the question as you can. \\
            (3) If the context is not clear, skip the citation. Make sure that every question is of the highest quality, as these questions will be used for important work concerning information extraction from the scientific literature. \\
            (4) The answer should ONLY contain the citation key. \\ \\ \\

            TEXT: We argue there are two underlying motivations for the ad text generation task, especially for product descriptions. Application-wise, the utility is to improve the seller experience for e-commerce services when registering a new product. The generated descriptions can reduce the need for manual data entry, and potentially improve sales due to better descriptions (in terms of attractiveness, structure, and persuasiveness). Research-wise, ad text generation is an under-studied task, and arguably a good proxy for persuasive text generation \{cite\_016\}\{cite\_017\}\{cite\_018\}\{cite\_019\}. \\ \\

            PAPER TITLES: \\
            - cite\_016: Is this post persuasive? ranking argumentative comments in online forum \\
            - cite\_017: On the role of discourse relations in persuasive texts \\
            - cite\_018: Measuring online debaters' persuasive skill from text over time \\
            - cite\_019: Analyzing the Persuasive Effect of Style in News Editorial Argumentation \\
            - cite\_020: Persuaide! an adaptive persuasive text generation system for fashion domain \\
            - cite\_021: A statistical framework for product description generation \\
            - cite\_022: SILVER: Generating persuasive Chinese product pitch \\
            \\

            QUESTION: I want to read some papers that try to study and quantify how persuasive text can be. I'm coming at this from an applications-perspective, as I'm interested in using these insights for product development. Could you give me a list of papers to read? \\
            ANSWER: cite\_016, cite\_017, cite\_018, cite\_019 \\ \\

            QUESTION: Has anyone looked at automating advertisement text generation specifically for fashion items? \\
            ANSWER: cite\_020 \\ \\

            QUESTION: I'm thinking about going into the computer-retail business and I'm wondering if it's possible to generate persuasive text to sell more computers? \\
            ANSWER: cite\_021 \\ \\

            QUESTION: Could you recommend some readings for articles that generate persuasive text in Chinese aimed at advertising? \\ 
            ANSWER: cite\_022 \\ \\

            \textbf{(Additional Demonstration Omitted)} \\\\
            TEXT: \{\{PARAGRAPH\}\} \\ \\

            PAPER TITLES: \\
            \{\{CITATIONS\}\} \\
            
        \bottomrule
    \end{tabular}
    \caption{
        The prompt used for generating questions from inline citations using GPT-4.
    }
    \label{tab:prompt_gen_question}
\end{table*}

\begin{table*}[h]
    \centering
    \small
    \begin{tabular}{>{\raggedright\arraybackslash\tt}p{0.95\textwidth}<{}}
        \toprule
            \vspace{-1em}

            \textbf{(system)} You are RankGPT, an intelligent assistant that can rank papers based on their relevancy to a research query. \\ \\
            
            \textbf{(user)} I will provide you with the abstracts of 100 papers, each indicated by number identifier []. \textbackslash nRank the papers based on their relevance to research question: \{\{query\}\}. \\ \\

            \textbf{(assistant)} Okay, please provide the papers. \\ \\

            \textbf{(user)} [1] Title: \{\{Title1\}\}\textbackslash n Abstract: \{\{Abstract1\}\} \\ \\

            \textbf{(assistant)} Received passage [1]. \\ \\

            \textbf{(user)} [2] Title: \{\{Title2\}\}\textbackslash n Abstract: \{\{Abstract2\}\} \\ \\

            \textbf{(assistant)} Received passage [2]. \\ \\

            \ldots \\ \\

            \textbf{(user)} [100] Title: \{\{Title100\}\}\textbackslash n Abstract: \{\{Abstract100\}\} \\ \\

            \textbf{(assistant)} Received passage [100]. \\ \\

            \textbf{(user)} Search Query: \{\{query\}\}. \textbackslash nRank the 100 papers above based on their relevance to the research query. The papers should be listed in descending order using identifiers. The most relevant papers should be listed first. The output format should be [] > [], e.g., [1] > [2]. Only respond with the ranking results, do not say any words or explain. \\

        \bottomrule
    \end{tabular}
    \caption{
        The prompt used for reranking retrieved documents  using GPT-4o (adapted from \citealp{Sun2023IsCG}). %
    }
    \label{tab:prompt_rerank}
\end{table*}

\begin{table*}[h]
    \centering
    \small
    \begin{tabular}{lcccccccc}
        \toprule
        &\multicolumn{3}{c}{\tf{Broad}} & \multicolumn{3}{c}{\tf{Specific}}  & \multirow{2}{*}{\tf{Total \#Q}} \\
        \cmidrule(lr){2-4} \cmidrule(lr){5-7} 
        & \#Q & Avg. L & Overlap & \#Q & Avg. L & Overlap & \\
        \midrule
        \multicolumn{8}{c}{\tf{\Inlineq{} Questions}} \\
        \midrule
        \tf{ACL-sourced} & 32 & 24.8 & 0.33 & 66 & 26.0 & 0.35 & 98 \\
        \tf{Non-ACL-sourced} & 88 & 19.1 & 0.33 & 165 & 20.5 & 0.34 & 253 \\
        \midrule
        \multicolumn{8}{c}{\tf{\Authorq{} Questions}} \\
        \midrule
        \tf{ACL 2023} & 25 & 14.5 & 0.41 & 130 & 18.1 & 0.42 & 155 \\
        \tf{ICLR 2024} & 10 & 19.0 & 0.49 & 81 & 17.6 & 0.45 & 91 \\
        \bottomrule
    \end{tabular}
    \caption{Detailed statistics for \ours{}.}
    \label{table:detailed_stats}
\end{table*}

\end{document}